\documentclass{elsart}
\usepackage{graphicx}
\begin{document}
\begin{frontmatter}
\title{Muon Flux at the Geographical South Pole}

\author[Bartol]{X.Bai \thanksref{bai}} 
\thanks[bai]{Corresponding author. Tel.: (302)831-8165;
E-mail address: bai@bartol.udel.edu.}
\author[Bartol]{T.K.Gaisser}
\author[Madison]{A.Karle}
\author[Madison]{K.Rawlins \thanksref{kath}}
\thanks[kath]{present address: University of Alaska Anchorage, 
3321 Providence Dr., Anchorage, AK 99508}
\author[Bartol]{G.M.Spiczak \thanksref{glenn}}
\thanks[glenn]{present address: Department of Physics,
 University of Wisconsin-River Falls, River Falls, WI 54022}
\author[Bartol]{Todor Stanev}
\address[Bartol]{Bartol Research Institute and Department of 
Physics and Astronomy, University of Delaware, Newark, DE 19716, USA}
\address[Madison]{Department of Physics, University of Wisconsin, Madison,
 WI 53706, USA}

\begin{abstract}
The muon flux at the South-Pole was measured for five zenith
angles, $0^{\circ}$, $15^{\circ}$, $35^{\circ}$, $82.13^{\circ}$ and
$85.15^{\circ}$ with a scintillator muon telescope incorporating ice 
Cherenkov tank detectors as the absorber. We compare the measurements 
with other data and with calculations.  
\end{abstract}
\end{frontmatter}

\section{Introduction and motivation}
The IceCube neutrino telescope~\cite{IceCube} under construction 
at the South Pole includes a surface array, IceTop, for calibration 
and tagging of cosmic-ray induced background. Together, the surface 
array and neutrino telescope constitute a three-dimensional air 
shower array that will also be used to study the cosmic-ray spectrum 
from 300 TeV to 1 EeV~\cite{ga03}. IceTop consists of ice Cherenkov
detectors with two tanks per station near the top of each IceCube 
string. The surface detector, when complete, will form a 
kilometer-square air shower array with nearest neighbor spacing 
between stations of approximately 125 m. The tanks contain 
clear ice with a surface area of 2.4 m$^2$ and depth of 0.9 m, 
and each is instrumented with two IceCube digital optical 
modules (DOMs) partially embedded in the ice and facing down.
Four two-tank stations with a total of 16 DOMs have been taking 
data since they were deployed in the 2004-2005 austral summer 
season ~\cite{ga05}.

Water Cherenkov detectors were used as the primary detector
elements in the Haverah Park surface array~\cite{av03} and they
make up the ground array of the Auger Project in Argentina~\cite{au05}. 
Following an early test at the South Pole~\cite{barw93},
development of a frozen Cherenkov detector as an element of 
an air shower array was undertaken in connection with the 
South Pole Air Shower Experiment (SPASE)~\cite{ba01}. Use 
of ice tanks as detectors for IceTop requires understanding 
the background radiation and in particular the muon fluxes 
at the South Pole. As in the Auger experiment~\cite{au05}, 
the pulse-charge distribution of single tank hits with its 
characteristic muon peak will be used for detector calibration 
and monitoring. In normal data taking muons will not be 
tagged by a muon telescope, so their fluxes must be well 
understood as a function of zenith angle on-site. In addition, 
muons are one of the major components of air showers,
becoming especially important for showers at large zenith 
angle. Therefore detailed study of detector response to muons
is needed to design the data acquisition system and interpret 
waveforms generated by air showers.

Measurements of the muon flux are also of intrinsic 
interest, for example, in connection with understanding 
the production spectra of atmospheric neutrinos. Although 
many muon flux measurements have been made throughout 
history (see the review by P.K.F~Grieder~\cite{gr01}), 
most of them were done for vertical or nearly vertical 
muons. Only a few experiments measured nearly horizontal 
muons above a few GeV, for example at 3220 m a.s.l~\cite{sh79} 
and 3250 m a.s.l~\cite{as75,as83}. In this work, using the 
test tanks deployed at the South-Pole as the absorber, we 
made a controlled and precise measurement of the flux of 
muons with minimum kinetic energy of several hundred MeV 
to about one GeV and zenith angle from vertical to 
nearly horizontal at the South Pole ($\sim700~g/cm^{2}$, or
$\sim3100$ m a.s.l.). The details of the setup are given in 
\S2 and the analysis and results of our measurements in \S3.
To give a systematic view of the muon flux at this altitude, 
some data from other measurements are collected in \S4.  
We conclude in \S5 with a comparison to calculations.

\section{Details of the setup}
Most muons at the surface of the Earth are produced high 
in the atmosphere from the decay of charged pions generated 
in the interactions of high energy cosmic rays with atmosphere 
nuclei. The muon flux varies with altitude and has a strong 
zenith angle dependence, reflecting the convolution of their 
production spectrum with their energy loss and decay.  
At large zenith angles the muon flux becomes very small,
as most of the muons generated in the atmosphere decay 
before reaching the detector. As a consequence, the 
experimental setups for measuring the nearly vertical and 
nearly horizontal muon fluxes are different. Near the vertical 
a simple coincidence requirement is sufficient, whereas 
the measurement at large zenith angles requires more 
complex trigger and analysis procedures because of the 
low event rate.

\begin{figure}[htb]
\begin{center}
\vspace*{2.0mm} 
\includegraphics[width=8cm,height=6cm]{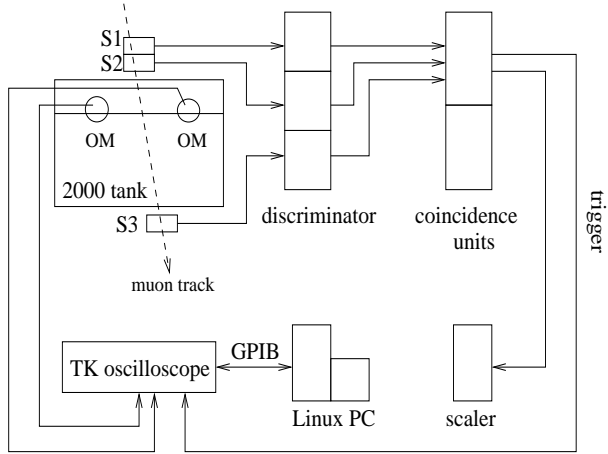} 
\caption
{A sketch of the ice Cherenkov detector (as the absorber), 
the muon telescope scintillators and the electronics in 
the flux measurement of \protect$0^{\circ}$, 
\protect$15^{\circ}$ and \protect$30^{\circ}$ muons. 
Three fold coincidence was used to effectively eliminate 
the accidental coincidence rate and other background 
triggers that may form a two fold coincidence by chance. 
\label{f1}
} 
\end{center}
\end{figure}

We show in Fig.~\ref{f1} the configuration used for measuring
the flux of nearly vertical muons. From December 10 to 24, 2000, 
the integral muon flux was measured at zenith angles of $0^{\circ}$,
$15^{\circ}$ and $30^{\circ}$. The absorber was a small ice Cherenkov
detector installed at the South Pole inside the boundary of the
SPASE-2 array~\cite{di00}. It is a cylindrical polyethylene tank
of area of $1.14~m^{2}$ and height of 1.24 m. The inside of the 
tank is lined with white, diffusely-reflecting Tyvek (type 1025D). 
After filling with the South-Pole station drinking water, two 
analog AMANDA optical modules (OMs)~\cite{an01}, with a separation 
of 0.51 m, were mounted facing down symmetrically off-center 
with their photo cathode region completely submerged. After 
the tank was frozen, the resulting ice depth was 0.99 m. A 
muon telescope consisting of three $0.2~m^2$ scintillators, 
two stacked on top of the tank and one underneath, identifies 
penetrating muons by requiring a coincidence among S1, S2 and 
S3 within 50 ns. Electric pulses from the two OMs and the 
three scintillators were transmitted through $\sim100$ m of 
twisted pair and RG-8 coaxial cables, respectively, to the 
central SPASE-2 building that houses the electronics and data 
acquisition system. Three scintillator signals first enter a
Phillips 711 discriminator, then go to a Phillips 755 logic unit. 
Both the singles rate and the coincidence rate were measured 
by a Jorway 1880B scaler. A 1 GHz Tektronix digital oscilloscope 
and a Linux PC were used to digitize and read out the waveforms 
from the two OMs through PCI-GPIB. 

\begin{figure}[ht]
\begin{center}
\vspace*{2.0mm} 
\includegraphics[width=10cm,height=3cm]{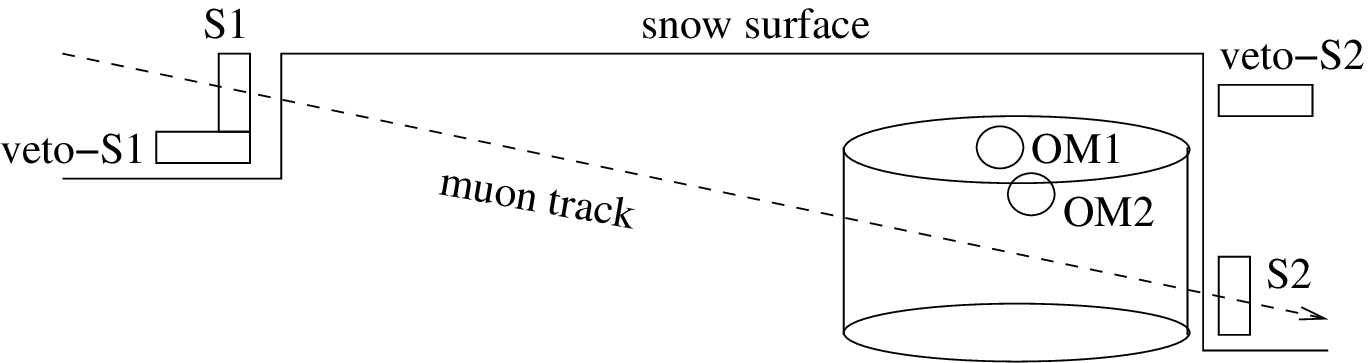}
\includegraphics[width=11cm,height=8cm]{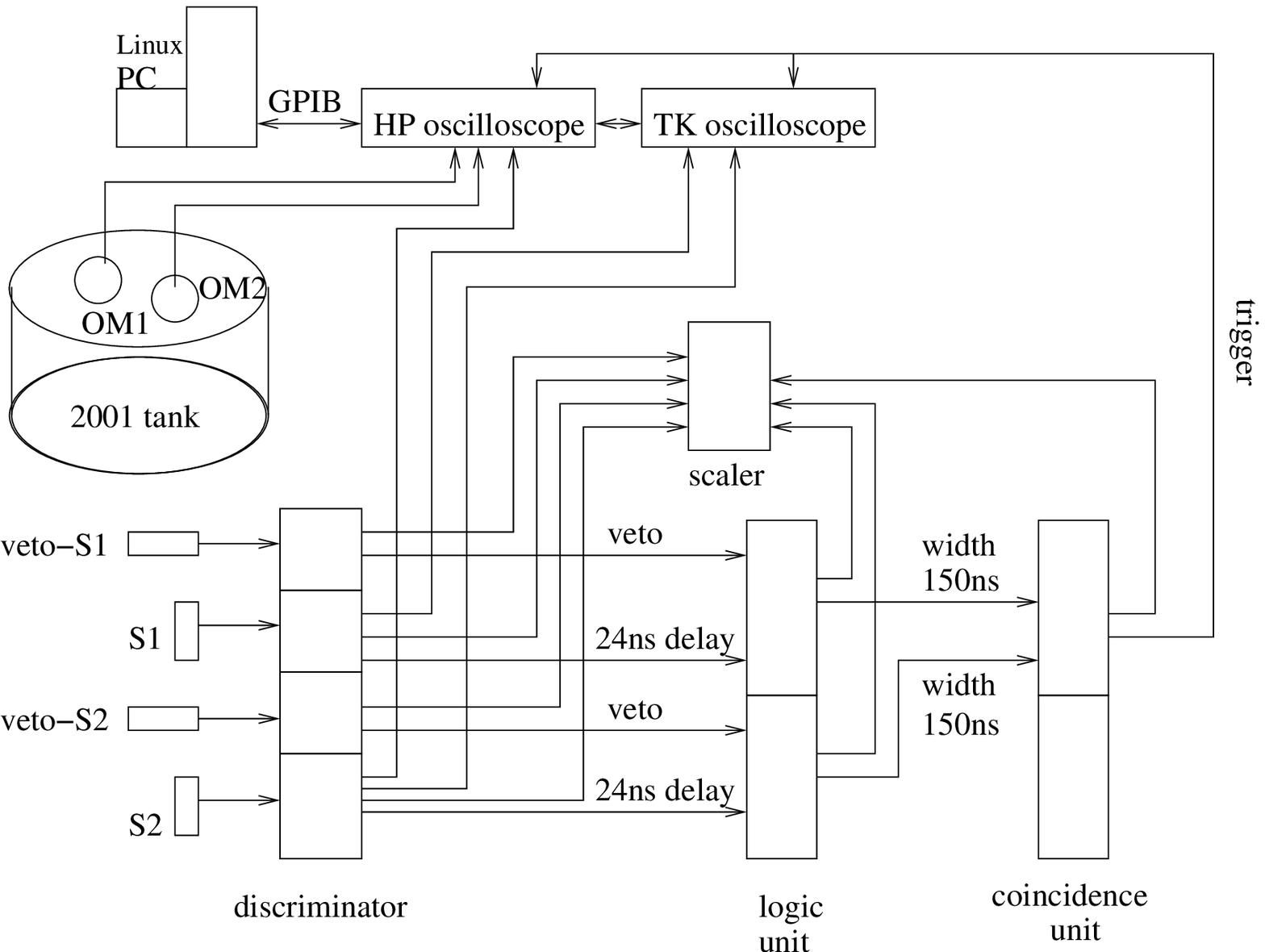}
\caption{A sketch of the detector layout (top) and 
the electronics (bottom) in the flux measurement of 
\protect$82.13^{\circ}$ and \protect$85.15^{\circ}$  
muons. Scintillators $S1$ and $S2$ were used to form 
a muon telescope with veto-S1 and veto-S2 in 
anti-coincidence. 
\label{f2}
}
\end{center}
\end{figure}

The configuration used for the measurement of horizontal 
muons is shown in Fig.~\ref{f2}. Because of the low rate of 
horizontal muons the coincidence rate of two scintillators 
forming a simple muon telescope is dominated by triggers of 
local small air showers. Reducing the background requires 
special attention to the detector setup and to the data 
taking. First, as shown in the setup sketched in 
Fig.~\ref{f2}, two scintillators marked 'veto-S1' and 
'veto-S2' were used in anti-coincidence with S1 and S2 
respectively. Second, in addition to the waveforms from
the two OMs in the tank, the trigger pulse and the traces 
of the discriminated pulses from the two muon telescope 
scintillators (S1 and S2 on the figure) were recorded by 
two digital oscilloscopes to provide the relative timing 
information among these pulses. The cuts used to filter 
the data are described in the next section. The singles 
rate of each scintillator and the coincidence rate with 
and without anti-coincidence were monitored by a scaler 
that has a negligible dead time at the working rates. The 
rates measured by the scalars were then used to correct 
for the dead time in the signal digitizing and reading system.

\begin{figure}[ht]
\begin{center}
\vspace*{2.0mm} 
\includegraphics[width=12cm,height=8cm]{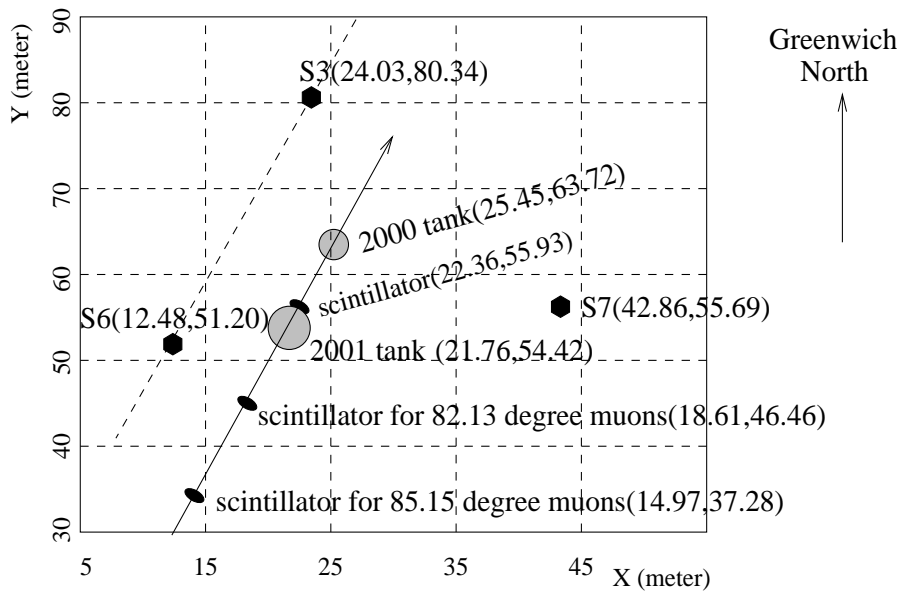}
\caption{The coordinates of the detectors in the 
nearly horizontal muon measurement. S3, S6 and S7 
are three of the SPASE-2 stations. Coordinate (X,Y) 
in meters are given in the SPASE-2 reference frame. 
\label{f3}
}
\end{center}
\end{figure}

The data for muon flux measurement at zenith angle 
$82.13^{\circ}$ and $85.15^{\circ}$ were taken from 
December 13 to 21 in 2002, during which some time was 
spent to rearrange the cables and detectors in the 
snow and to survey the detector positions. In this 
measurement, the tank was a larger cylindrical test 
tank ($3.94~m^{2}\times 1.23~m$) installed inside the 
boundary of the SPASE-2 array in 2001. The preparation 
and installation of this tank were similar to the 
smaller tank. The two OMs are positioned symmetrically 
about the center with a separation of 1.07 m. The 
ice depth in the tank is 1.06 m. Fig.~\ref{f3} 
shows a plan view of the coordinates of the tank and 
scintillators within the SPASE array. 

\section{Data analysis and muon flux at the South Pole}
\subsection{Nearly vertical muons}
The flux of penetrating muons at three near vertical
zenith angles is summarized in Table~\ref{tab:1}. Only 
the statistical errors are given. Column three gives the 
acceptances of the muon telescope at these three zenith 
angles. The acceptances were obtained by a Monte Carlo 
simulation that requires $5.0\times 10^4$ particles to 
pass through all three scintillators for each case. The 
last two columns show respectively the integral fluxes 
and the minimum kinetic energy (and momentum) required 
for muons to pass through all the three scintillators.
The flux is obtained as the measured rate of events 
divided by the acceptance.

\begin{table}[t]
\begin{center}
\caption{Rates and fluxes of muons near the vertical at the 
South Pole. The muon threshold kinetic energy was calculated 
from the table of the continuous-slowing-down-approximation 
(CSDA) muon range in water\protect~\cite{gro01}. 
The ice density used in the calculation is 
$0.92$~g$\cdot$cm$^{-2}$.
\label{tab:1}}
\vspace*{0.4cm} 
\begin{tabular}{ccccc}
\hline
zenith(min,max) &   rate  &  acceptance   & integrated flux  & $E_{min}(p_{min})$  \\
 $(^{\circ})$   &   $(Hz)$  &  $(cm^{2}\cdot sr)$  & 
 $(cm^{2}\cdot s\cdot sr)^{-1}$  &  $MeV(MeV/c)$    \\
\hline
$0.0 (0,16.5)$     & $2.22\pm0.01$  &  
$1.26\times10^{2}$  &  $(1.76\pm0.01)\times10^{-2}$ &  246.0(335.4)  \\
$15.0(0.4,28.6)$   & $1.80\pm0.01$  &  
$1.07\times10^{2}$  &  $(1.68\pm0.01)\times10^{-2}$ &  263.5(353.7)  \\
$35.0(25.8,44.2)$  & $0.72\pm0.01$  &  
$4.84\times10^{1}$  &  $(1.49\pm0.02)\times10^{-2}$ &  311.7(403.8)  \\
\hline
\end{tabular}
\end{center}
\end{table}

\subsection{Nearly horizontal muons}
\begin{figure}[ht]
\begin{center}
\vspace*{0.4cm} 
\includegraphics[width=11cm,height=12cm]{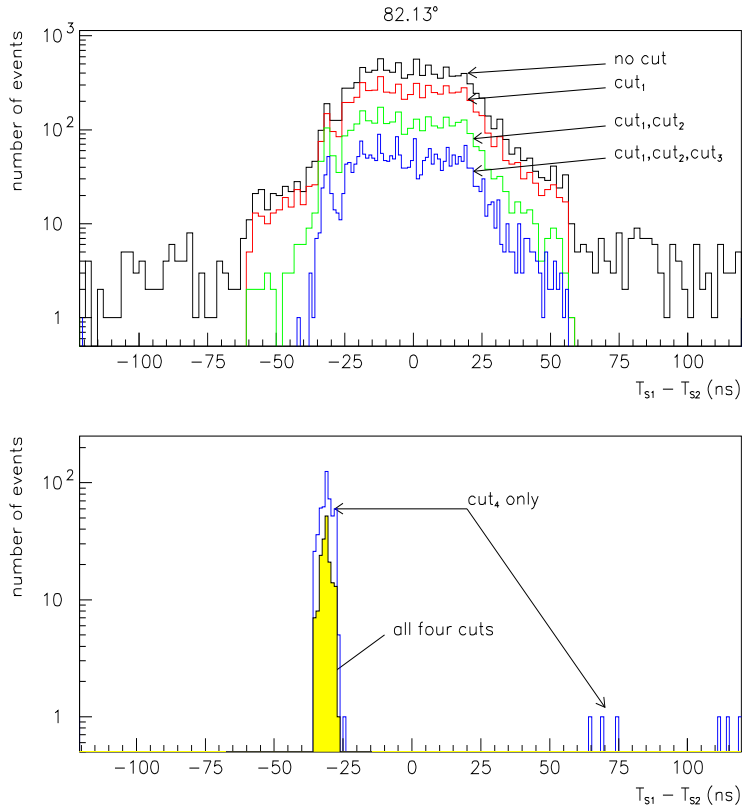} 
\caption{
The TOF ($T_{S1}\,-\,T_{S2}$) spectra of triggered events 
in \protect$\theta=82.13^{\circ}$ muon measurement. The cuts 
used for selecting the events under each histogram are marked. 
See more details about the cuts in the text. \label{f4}
}
\end{center}
\end{figure}

\begin{figure}[ht]
\begin{center}
\vspace*{0.4cm} 
\includegraphics[width=11cm,height=12cm]{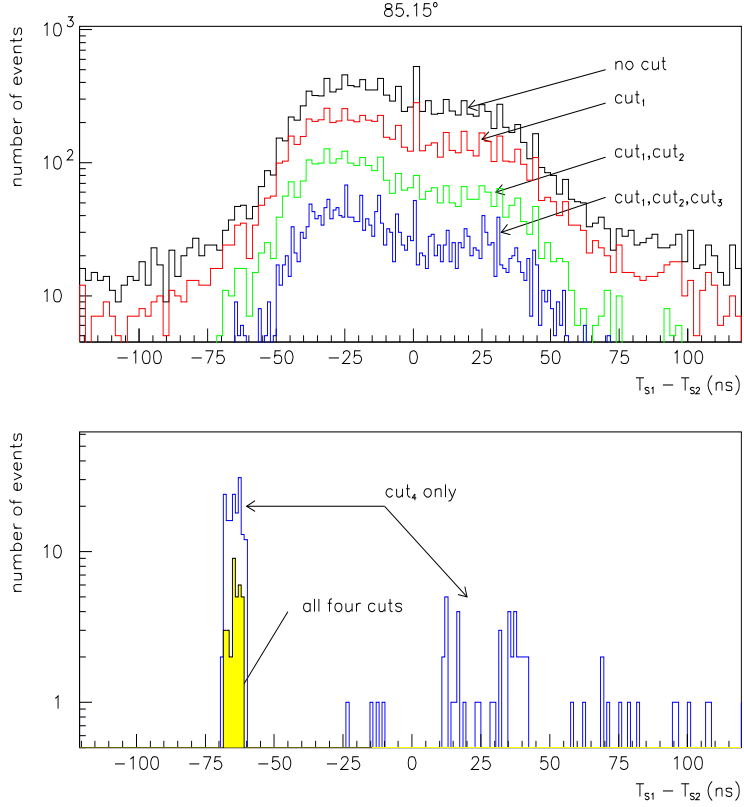} 
\caption{
The TOF spectrum of triggered events in $\theta=85.15^{\circ}$ 
muons measurement.
\label{f5}
} 
\end{center}
\end{figure}

Because of the low horizontal muon flux and the relatively
high rate of local small air showers, the coincidence rate 
between scintillator $S1$ and $S2$ on Fig.~\ref{f2} is 
dominated by background even when the two veto scintillators
are included in the trigger. Figures~\ref{f4} and \ref{f5} 
show the distributions of time of flight (TOF) as recorded 
by scintillators $S1$ and $S2$, ($TOF = T_{S1}\,-\,T_{S2}$), 
with no cuts (the top histogram) and after various cuts 
have been applied to the data. The physical distance from 
$S1$ to $S2$ corresponds to a time differences of 
$T_{S1}\,-\,T_{S2}\,=\,-30.5$~ns for the setup at 
$82.1^\circ$ and to $-63.5$~ns for the larger angle.  
In the top histogram (without cuts) there is little or 
no sign of a peak at the expected times. Therefore, 
simply counting all events on a TOF window would greatly 
overestimate the true horizontal muon flux. The following 
series of cuts was applied to reduce the background:

\begin{itemize}
\item $cut_{1}$: For each triggered event, scanning the 
waveforms recorded by the digital oscilloscopes and 
choosing those with exactly one hit in the two triggering 
scintillators and two OMs in a time window of 500 ns. 

\item $cut_{2}$: Choosing events in which the peaks of 
the pulses in the two OMs in the tank are close in time 
$abs(T_{OM1}-T_{OM2})\leq 14~ns$. The 14 ns is related 
to the slow rise time and the signal fluctuations in 
the tank. This cut selects those events of which the 
Cherenkov photons generated in the tank give rise to 
the signals in the two OMs evenly and simultaneously. 
The first two cuts are both designed to reduce the number 
of events caused by showers with several particles spread 
in time. In addition to single particle events, however,
events with a few particles arriving within a time window 
less than the detector resolving times can still pass 
these cuts.

\item $cut_{3}$: Choosing events in which 
$abs(T_{OM1}\,-\,T_{S2}\,-\,234~ns)\leq 16~ns$. 
Here, $T_{S2}$ is the time of the discriminated pulse
from the trigger scintillator $S2$. The average cable 
delay between the $OM1$ signal and the signal from 
scintillator $S2$ recorded on the oscilloscope is 
$234$~ns. It was determined by a system calibration 
with vertical muons in which we used the same cable, 
the same trigger threshold, etc. The $16~ns$ is one 
half of the rise time of the OM pulse, the fluctuation 
in the tank signal as seen in $OM1$ and the fluctuation 
in the signal from scintillator $S2$. This cut gets 
rid of those triggers in which the pulse generated 
in $S2$, which is located next to the tank, is too 
far in time from the signal in the tank. 

\item $cut_{4}$: Choosing events in which
$abs(T_{S1}\,+\,T_{TOF}\,+\,T_{delay})\leq\delta(t)$.
Here, $T_{S1}$ is the time (on the oscilloscope trace) 
of the discriminated pulse from scintillator S1. In the 
setup, it is a negative number (i.e. to the left of the 
trigger moment on the oscilloscope trace) for single 
muon event. $T_{TOF}$ is the muon time-of-flight from 
S1 to S2, which is 30.5 ns and 63.5 ns for muons of 
$\theta=82.13^{\circ}$ and $\theta=85.15^{\circ}$ in 
this setup. $T_{delay}$ is the 50 ns trigger delay set 
in the digital oscilloscope. For single muon events, 
$abs(T_{S1}\,+\,T_{TOF}\,+\,T_{delay})$ is the time 
jitter relative to the DAQ trigger moment which is 
$0$~ns on the oscilloscope trace. In the analysis, 
$\delta(t)$ was chosen 4.5 ns, which corresponds to 
the time jitter in the scintillator and the electronics.
\end{itemize}

In Fig.~\ref{f4} and \ref{f5}, from top to bottom, the
$S1$ to $S2$ TOF spectrum are shown for triggered events 
without any cut, and events left after the application of
cuts 1 to 4. The following features on the TOF spectra are
noted:
\begin{itemize}
\item[(a)] The events left after the application of the 
first three cuts start to show a clear peak at the correct 
TOF position on the spectrum although none of these cuts 
makes use of the time-of-flight from $S1$ to $S2$;
\item[(b)] Although most of the events that survive the
$cut_{4}$ (and no other cuts) distribute themselves 
around the expected muon flight time on the TOF spectrum, 
several of them still fall outside the expected time window
in Fig.~\ref{f4}, and more in Fig.~\ref{f5};
\item[(c)] All the events left after all four cuts show 
themselves at the expected position as on an ideal TOF 
spectrum of single muon events;
\item[(d)] The events in the peak with all four cuts applied 
correspond to those events that make up the peak on the 
distribution with cuts 1, 2 and 3.  
\end{itemize}

We therefore conclude that the events left after application
of all four cuts are true horizontal muon events. It is 
interesting to compare their waveforms with vertical muon 
waveforms as seen by the two OMs in the same tank. We note 
two features in the data: 
\begin{enumerate}
\item The average amplitude of these pulses is about 2 times
as big as that of the vertical muons. 
\item The average rise time of the these pulses in the tank is
3ns longer than that of vertical muon pulses.  The fall time of 
the average waveform is nearly the same as that of the average 
waveform of vertical muons.
\end{enumerate} 

A GEANT-4 simulation of the tank response was carried out 
for both vertical muons and muons of zenith angle 
$82.13^{\circ}$ and $85.15^{\circ}$~\cite{ul03}. 
The two features listed above are well reproduced in the 
simulation. They are directly related to the longer track 
length of the near horizontal muons in the tank.

Taking the events left after all four cuts as the single muon 
events, we summarize the measurement in Table~\ref{tab:2}.

\begin{table}[ht] 
\caption{Summary of the horizontal muon measurement at the
South-Pole. The data taking time excludes dead time. The
systematic errors are from the uncertainty in the cut
parameters. They were obtained by a systematic comparison
of the S1 to S2 time difference spectra under four cuts
with different parameter values. The muon threshold kinetic
energy was calculated from the table of the continuous-
slowing-down-approximation (CSDA) muon range in water 
\protect~\cite{gro01}. The ice and the snow densities used
in the calculation are $0.92$~g$\cdot$cm$^{-2}$ and 
$0.4$~g$\cdot$cm$^{-2}$. The latter has an error of $\pm5\%$.
\label{tab:2}
}
\vspace*{0.4cm} 
\begin{tabular}{ccc}
\hline
zenith coverage $(^{\circ})$            &     $82.13\pm2.97$           &    $85.15\pm1.44$  \\
time-of-flight $(ns)$                &     $30.5$                   &    $63.5$                   \\
number of events                     &     $173$                    &    $33$                     \\
acceptance $(cm^{2}\cdot sr)$        &     $4.6881$                 &    $1.0944$                 \\
data taking time(sec)                &     $2.1227\times 10^{5}$    &    $2.5600\times 10^{5}$    \\
$E_{min}(MeV)(p_{min}(MeV/c))$       &     $899\pm45(999\pm50)$     &    $1752\pm87(1855\pm93)$   \\
statistical errors                   &     $\pm7.6\%$               &    $\pm17.4\%$              \\
systematical errors                  &     $^{+14\%}_{-13\%}$    &    $^{+3\%}_{-12\%}$        \\
flux $(cm^{2}\cdot s\cdot sr)^{-1}$  &     $(1.74^{+0.24}_{-0.22}\pm0.13)\times 10^{-4}$      
                                     &     $(1.18^{+0.04}_{-0.14}\pm0.21)\times 10^{-4}$       \\
\hline
\end{tabular}
\end{table}
\vspace*{0.4cm} 

\section{Muon fluxes by some other experiments at an 
altitude close to the South-Pole} 

There are several measurements of the vertical muon flux 
at high altitude that are performed at different muon 
threshold energies, geomagnetic locations and solar epochs. 
At 3220 m a.s.l, 
the vertical intensity of muons above $2\;GeV$ was measured 
with GM hodoscope by Shen and Chiang~\cite{sh79}. The reported 
flux is 
$(4.9\pm0.2)\times 10^{-3} (cm^{2}\cdot s\cdot sr)^{-1}$. 
For muons of energy $\geq162.9MeV$, the vertical integral 
intensity at the South-Pole altitude is 
$(2.61\pm0.01)\times 10^{-2} (cm^{2}\cdot s\cdot sr)^{-1}$ 
according to the curve on figure 2.130 in reference~\cite{gr01},
which was made from the measurement by Blokh et al.~\cite{bl77}. 
Vertical muons at 2960m a.s.l. 
and 3250m a.s.l. 
were measured by Allkofer and Tr\"{u}mper~\cite{al64} 
and Kocharian et al.~\cite{ko56}. By integrating the 
differential momentum spectrum on figure 2.137 in 
reference~\cite{gr01}, the integral muon intensities 
were found to be 
$I_{2960m}(\geq 203\;MeV/c)=1.16\times 10^{-2} (cm^{2}\cdot s\cdot sr)^{-1}$ 
and 
$I_{3250m}(\geq414\;MeV/c)=1.58\times 10^{-2} (cm^{2}\cdot s\cdot sr)^{-1}$. 
Some of the differences between these results can be explained 
by the different altitudes and muon threshold energies.
The difference in altitude between Ref.~\cite{al64} and 
South Pole corresponds to about 30 g/cm$^2$, which 
increases the energy threshold by 60 MeV. In addition, 
there is also a decrease of the muon flux because of 
muon decay. About 15\% of the muons at threshold decay 
in 300 m. These effects, as well as the solar epochs, do not
explain all the differences among the quoted measurements.
Our result is in the middle of the scattered points 
obtained by those previous work and agrees the best with 
the result of Ref.~\cite{ko56}.

 Few experiments have been done for low energy nearly horizontal
 muons with a small zenith bin. For an angular interval 
 from $78.4^{\circ}$ to $90^{\circ}$, corresponding to a 
 mean of $86.2^{\circ}$, the intensity at 3220 m a.s.l
 was reported to be~\cite{sh79}:  
 $I_{h}(\geq2\;GeV)=(9.68\pm0.32)\times 10^{-5}
 (cm^{2}\cdot s\cdot sr)^{-1}$.
 Because of the large angular range and the steep dependence 
 of the muon flux on the cosine of the zenith angle
 this number which is lower than the current measurement 
 is difficult to interpret.  

 Asatiani et al.~\cite{as75} carried out a spectral
 measurement at 3250 m a.s.l with a large magnetic 
 spectrometer. The muon differential spectrum covers
 the range from $10\;GeV/c$ to $2000\;GeV/c$. The muon
 incident angle is from $80^{\circ}$ to $90^{\circ}$
 with the mean zenith angle $84^{\circ}$.
 The presented spectrum was normalized to the
 calculation by Ashton et al.~\cite{as66} at $50\;GeV$
 for the zenith angle $84^{\circ}$. Integrating over 
 the differential energy spectrum in figure 2.139 in 
 reference~\cite{gr01} gives an approximate intensity of 
 $(8.85\pm0.02)\times 10^{-5} (cm^{2}\cdot s\cdot sr)^{-1}$.

With the same instrument, Asatiani et al.~\cite{as83} 
reported the differential muon spectra at narrow zenith
angle intervals covering from $80^{\circ}$ to $88^{\circ}$.
The integral muon flux in units of $(cm^{2}\cdot s\cdot sr)^{-1}$ 
above a given energy threshold for four zenith bins are: 
$80^{\circ}-82^{\circ}$, $I_{>16.5GeV}= 1.27\times 10^{-4} $;
$82^{\circ}-84^{\circ}$, $I_{>16.5GeV}= 1.01\times 10^{-4}$;
$84^{\circ}-86^{\circ}$, $I_{>25.8GeV}= 5.60\times 10^{-5}$;
$86^{\circ}-88^{\circ}$, $I_{>14.8GeV}= 4.40\times 10^{-5}$. 

 Because of the finer binning in the zenith angle we can 
 correct these numbers for the difference in muon threshold
 and compare to the current measurement. The correction
 was done using the calculation of Ref.~\cite{ag96}
 for the fluxes of muons at $\cos \theta$ = 0.25, 0.15
 and 0.05 at the sea level. The muon fluxes from this 
 calculation were interpolated for the average 
 $\cos \theta$ and energy threshold in the measurement 
 and compared to the fluxes above 2 GeV energy threshold. 
 The scaling factors for the four points were found to 
 be 1.84, 1.60, 1.77 and 1.22 respectively. The data of 
 Ref.~\cite{as83}, after being scaled to muon energy 
threshold of 2 GeV, are shown with open squares in 
Fig.~\ref{f6}. The agreement is excellent. 

\section{Comparison with simulation} 
The measurements were compared with simulations. For the 
vertical and near vertical muons, a Monte Carlo  
simulation was carried out to calculate the fluxes expected 
for the average atmosphere depth of $702~g/cm^{2}$ during 
the muon measurement. The simulation used TARGET2.1~\cite{en01} 
and the cosmic ray flux of reference~\cite{ag96} 
for the epoch of solar maximum. The muon fluxes 
of $1.80\times10^{-2}$, $1.73\times10^{-2}$ and 
$1.68\times10^{-2}$ $(cm^{2}\cdot s\cdot sr)^{-1}$ were 
predicted for the three angles. The measured muon fluxes 
at $0^{\circ}$ and $15^{\circ}$ are in excellent agreement 
with the prediction, while at $35^{\circ}$ the flux is 
lower by $12\%$. 

\begin{figure}[htb]
\begin{center}
\vspace*{0.4cm} 
\includegraphics[width=11cm]{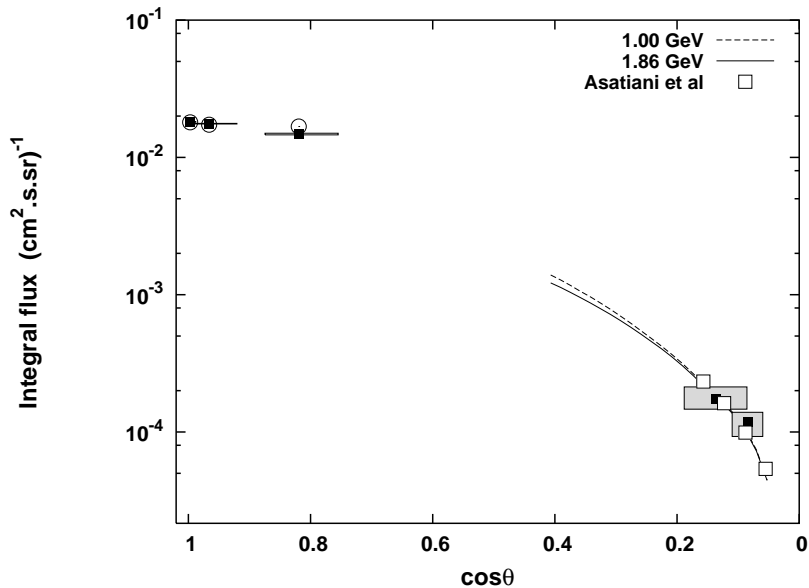} 
\caption{The measured integral flux of muons above the energy 
 thresholds specified in Tables~\protect\ref{tab:1} 
 and~\protect\ref{tab:2} are shown with full squares versus 
 the cosine of the zenith angle. The shaded boxes around 
 data points show the errors in the fluxes and the zenith 
 angular coverage in the measurement. The open circles show 
 the calculation for vertical and near vertical muons at 
 the South Pole. The lines show the angular dependence of the 
 predicted integral fluxes above the two energy thresholds. 
 The open squares are the measurements of Ref.~\cite{as83} 
 after being scaled to a threshold of 2 GeV. 
\label{f6} }
\end{center}
\end{figure}

The fluxes of nearly horizontal muons by this work are 
compared to the the simulation of Ref.~\cite{ag96}. The 
current measurements are shown with solid squares in 
Fig.~\ref{f6}. The predicted angular dependence of muons 
with energy above 1 (1.86) GeV from Ref.~\cite{ag96} are 
shown with a dashed (solid) line. One can see that the 
expected fluxes for the two muon energy thresholds at 
$\cos \theta$ less than 0.2 become the same. This is 
because the horizontal muon energy spectrum has very 
few muons in this energy range. The good agreement 
between the measurements and the calculations also 
confirms that accounting for difference in the altitude 
of the measurements is not important for nearly horizontal 
muons.

\section{Summary}
Using the prototype IceTop ice Cherenkov detector as the 
absorber, data has been collected for muons in five 
directions from vertical and nearly vertical to nearly 
horizontal. To select the rare nearly horizontal muon 
events, four cuts were developed in the data
analysis. The procedure for selecting horizontal muons 
was crosschecked by a GEANT-4 simulation of the muon 
waveforms seen in the OMs. 

 The integral muon flux of muons at $0^{\circ}$ and $15^{\circ}$
 agrees very well with the simulation, which accounts for the
 atmospheric depth and the specific atmospheric density profile
 above the South Pole. The simulation of Ref.~\cite{ag96} is 
 not done for the South Pole location, however this is not 
 important for the high energy muons that survive at very 
 large zenith angles. 

\begin{ack}

 The work was supported by the Office of Polar Programs of 
 the U.S. National Science Foundation under grants OPP-998081 
 and OPP-0236449. 
 The authors gratefully acknowledge the support from the U.S. 
 Amundsen-Scott South Pole station. The assistance from 
 Jerry Poirier during the detector deployment in 2000 and 
 2001, the GEANT-4 simulation done by Ralf Ulrich, and the 
 acceptance calculation in the near vertical muon measurement 
 by Zhe Ma are highly appreciated as well.
\end{ack}


\begin{thebibliography}{99}

\bibitem{IceCube} J. Ahrens et al. 
               Astropart. Phys. {\bf 20}, 507(2003). See also 
               {\it Icecube Project Preliminary Design Document}, 
               {\it http://icecube.wisc.edu/$pub_{-}and_{-}doc$/9000-0010.curr.pdf } 
\bibitem{ga03} T.K.~Gaisser for the IceCube Collaboration, 
               "IceTop: the Surface Component of IceCube",
               {\it Proc. $28^{th}$ Int.Cos.Ray Conf.}, Tsukuba, Japan,
               {\bf HE1.5}, 1117(2003) and also T.~Stanev,
               "IceTop Status in 2004",
               Nucl. Phys. B (Proc. Suppl) 
               {\bf 145}, 327(2005) 
\bibitem{ga05} T.K.~Gaisser for the IceCube Collaboration, 
               "Air showers with IceCube: First Engineering Data" 
               {\it Proc. $29^{th}$ Int.Cos.Ray Conf.}, 2005, Pune, India. See also 
               astro-ph/0509330, pp60-63. 
\bibitem{av03} M.~Ave, J.~Knapp, J.~Lloyd-Evans, M.~Marchesini, \& A.A.~Watson, 
               Astropart.~Phys. {\bf 19}, 47(2003) 
\bibitem{au05} Auger Collaboration (S.~Ranchon, M.~Urban), 
               NIM-A{\bf 538}, 483(2005). See also 
               Auger Collaboration, J.~Abraham et al., NIM-A{\bf 523}, 50(2004)
\bibitem{barw93} S.Barwick and J.Beaman,
               {\it Proc. $23^{rd}$ Int.Cos.Ray Conf.}, Calgary, 
               {\bf 4}, 683(1993)
\bibitem{ba01} X.Bai et al.,
               {\it Proc. $27^{th}$ Int.Cos.Ray Conf.}, Hamburg, Germany,
               {\bf 3}, 981(2001)
\bibitem{gr01} Peter K.F.Grieder, 
               {\it "Cosmic Rays at Earth Researcher's Reference Manual and Data Book"}, 
               Elsevier, 2001.
\bibitem{sh79} Shen Chang-quan and Chiang In-lin 
               {\it $16^{th}$ Int.Cos.Ray Conf.}, Kyoto, Japan,
               {\bf 10}, 8(1979) 
\bibitem{as75} T.L.Asatiani et al. 
               {\it $14^{th}$ Int.Cos.Ray Conf.}, Munchen, Germany,
               {\bf 6}, 2024(1975)
\bibitem{as83} T.L.Asatiani et al. 
               {\it $18^{th}$ Int.Cos.Ray Conf.}, Bangalore, India,
               {\bf 7}, 47(1983)
\bibitem{di00} J.E.Dickinson et al. 
               NIM-A
               {\bf 440}, 95(2000) 
\bibitem{an01} E.Andres et al. 
               Nature 
               {\bf 410}, 441(2001)
\bibitem{gro01} D.E.Groom, et al. 
               {\it $http://panisse.lbl.gov/\sim deg/muon.html$} 
\bibitem{ul03} Ralf Ulrich 
               {\it private communication}, 2003
\bibitem{bl77} Y.L.Blokh, et al. 
               Nuovo Cimento 
               {\bf 37B}, 198(1977) 
\bibitem{al64} Allkofer, O.C., and J. Tr\"{u}mper, 
               Z. Naturforsch 
               {\bf 19A}, 1304(1964)
\bibitem{ko56} Kocharian, N.M. et al. 
               Soviet Phys. JETP 
               {\bf 3}, 350(1956) 
\bibitem{as66} F.Ashton et al. 
               Proc.Phys.Soc. 
               {\bf 87}, 79(1966)
\bibitem{ag96} Vivek Agrawal, T.K. Gaisser, Paolo Lipari, and Todor Stanev 
               Phys. Rev.
               {\bf D53}, 1314(1996) 
\bibitem{en01} R.~Engel, T.K.~Gaisser, P.~Lipari, and T. Stanev, 
               {\it Proc. $27^{th}$ Int.Cos.Ray Conf.}, Hamburg, Germany. 
               {\bf 4}, 1381(2001)
\end{thebibliography}
\end{document}